# STUDY OF NEW FNAL-NICADD EXTRUDED SCINTILLATOR AS ACTIVE MEDIA OF LARGE EMCAL OF ALICE AT LHC

O. A. GRACHOV[†]

*Department of Physics and Astronomy, Wayne State University,*
*Detroit, MI 48201, USA*

T. M. CORMIER

*Department of Physics and Astronomy, Wayne State University,*
*Detroit, MI 48201, USA*

A. PLA-DALMAU, A. BROSS

*Fermi National Accelerator Laboratory,*
*Batavia, Il 60510, USA*

V. RYKALIN

*Northern Illinois Center for Accelerator and Detector Development (NICADD),*
*Northern Illinois University, DeKalb, Il 60115, USA*

The current conceptual design of proposed Large EMCal of ALICE at LHC is based largely on the scintillating mega-tile/fiber technology implemented in CDF Endplug upgrade project and in both barrel and endcap electromagnetic calorimeters of the STAR. The cost of scintillating material leads us to the choice of extruded polystyrene based scintillator, which is available in new FNAL-NICADD facility. Result of optical measurements, such as light yield and light yield variation, show that it is possible to use this material as active media of Large EMCal of ALICE at LHC.

**Introduction**

The ALICE-USA collaboration proposes to build a large area electromagnetic calorimeter (Large EMCal). The main goal in the design of Large EMCal is to significantly extend the limit coverage of the PHOS calorimeter [1]. The Large EMCal will be central to ALICE's program for the study of hard processes.

The calorimeter will cover the major part of the ALICE central acceptance, see Figure 1. The full detector spans $\Delta\eta = \pm 0.7$ with azimuthal acceptance of $\Delta\phi = 120°$ and segmented into ~20.000 towers projective in $\phi$ and $\eta$ to the interaction vertex. Minimum size of each tower is approximately 50 mm by 50 mm. The chosen technology is a tile/fiber sampling technique with scintillator and absorber plates perpendicular to the incident particles. The calorimeter structure consists of 25 active 5-mm scintillator layers interspersed with 5-mm lead absorption plates. The constant size in $\Delta\phi$ and $\Delta\eta$ and projective nature of the towers means that there are several thousands different sizes geometries. The technology chosen for the construction of the Large EMCal is that of megatile production as used in the CDF Endplug upgrade project and in both endcap and barrel calorimeters of the STAR. [2, 3, 4].

---

[†]e-mail address: grachov@physics.wayne.edu (Oleg A.Grachov)

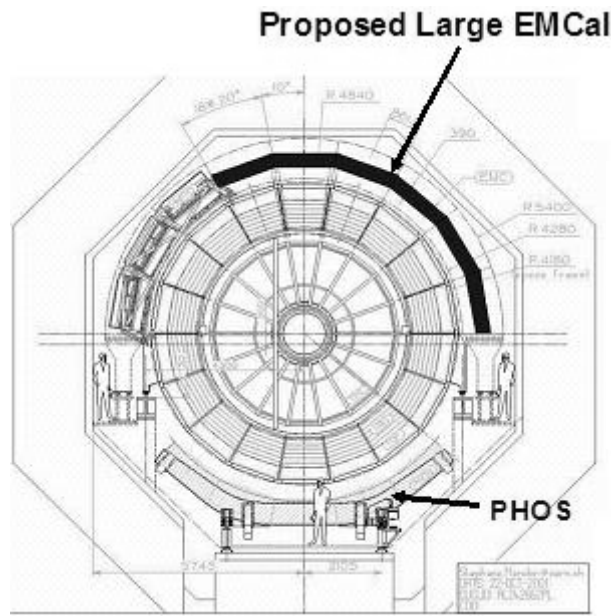

Figure 1. ALICE end view with proposed Large EMCal.

The Large EMCal required more than 15 tons of scintillating material. The thousands of different tile geometries make production technique such as injection molding far less attractive and become impractical because of the cost of the molds. Cast plastic scintillator sheets (at BICRON, Kuraray) may cost about $40-$70 per kg. The estimated price for extruded scintillator strips, which are become available at Fermilab, ranges between $5 and $8 per kg. [5]. These issues first lead us to the choice of extruded scintillator and drove the R&D into extruded material at FNAL-NICADD (Northern Illinois Center for Accelerator and Detector Development, Northern Illinois University) facility. One of the most important problems of optical system related to the scintillator is to achieve a specified minimum light yield and maintain high spatially uniformity, so that the resolution of calorimeter is not compromised by a lack of photostatistics or position dependence. Physics requirements lead us to require a light yield more than 2 photoelectrons per minimum ionizing particle (for full optical chain including photo detector) and uniformity of response over the surface of tile better than 3% RMS. Because the Kuraray scintillator has been used many times in large scale, long use-time applications and meet our requirements, we concentrated on comparison of Kuraray and the FNAL-NICADD scintillator. We have compared the light yield and light yield variation of individual tiles produced from Kuraray SCSN-81 [6] and FNAL-NICADD extruded scintillator.

**Measurements**

Kuraray SCSN-81 and new FNAL-NICADD scintillator of 5 mm thickness were used for production of tiles.

A first prototype production batch of scintillating strips was completed at FNAL in September 2003. This is blue – emitting scintillator with an absorption cut-off at ~ 400 nm (for a 1-cm path length) and an emission maximum at 420 nm. The optical characteristics of bulk material are the same as that of the MINOS extrusions [7].

For measurements we used tiles with dimensions 65 mm x 71 mm (maximum size, which required current design). A set of three identical tiles from each material has been produced. A peripheral sigma-groove was machined in each tile, where a Kuraray Y11 (200 ppm) MS WLS fiber diameter 0.83 mm was inserted. In order to exclude dependence of results from quality of fibers, we used one fiber. Length of fiber was 60 cm. The both end of WLS fiber were polished and than mirrored on one end. The other end of fiber was coupled to PMT (Hamamatsu R6094) with optical grease. The non-polished tile edges are painted white and tile was wrapped with light reflective material. All design parameters of WLS fiber read-out groove such as shape, depth, width, and distance from edges of tile were optimized for SCSN-81 scintillator during R&D phase for Barrel calorimeter of STAR at RHIC.

In order to measure the light yield and monitor light yield variation a specialized test setup has been design and build at Wayne State University, see Figure 2. The apparatus consist of DaVinci XYZ positioning system with bridge mounted Y-Z slide. Tile under test is mounted on the trigger counter. Trigger counter and PMT are mounted on the X slide. Cosmic ray counter is mounted below movable X slide. Radioactive source with collimator is mounted on the Y-Z slide. Combination of independent movements of X and Y-Z slides controlled by a personal computer allowing the light output of a tile to be scanned transversely as a function of position of radioactive source. The range of motion

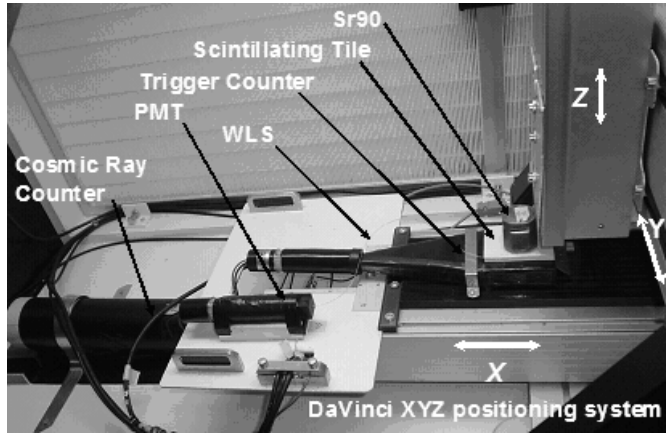

Figure 2. Test setup for light yield and light yield variation measurements.

is 200 mm in X – direction and 250 mm in Y – direction. The precision of motion is 0.1 mm.

The light yield from the tiles was measured using cosmic ray. Trigger was made with coincidence of two counters positioned behind the tested scintillating tile. Due to the difference in the area between system of trigger counters and scintillating tile under test there were empty triggers that permitted measuring of PED signal. Typically 3000 events were collected for each tile. To obtain the average light yield information, the spectra were fitted with a Gaussian function, see Figure 3.

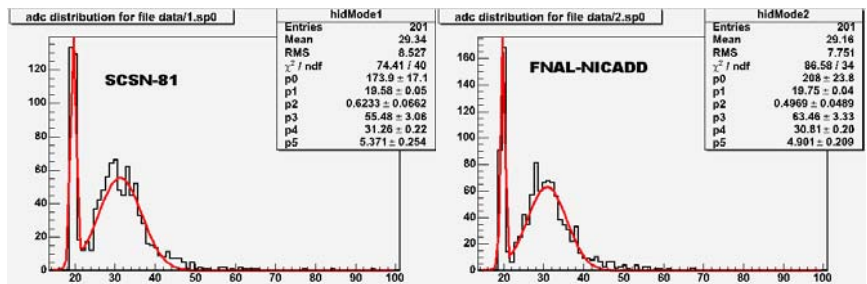

Figure 3. The amplitude spectrums for cosmic ray crossing scintillating tiles.

We estimated the absolute light yield using previous measurements of single photoelectron spectra. Tiles produced from two different scintillating materials always show the same light yield. The average light yield is 8 photoelectrons per mip; uncertainty of light yield is 5% due to insertion of fiber into read out groove. The uniformity of response has been measured using the collimated Sr90 beta ray radioactive source. We used lead collimator of 5-mm thick with hole of 2-mm diameter. Face of tiles was scanned with steps 8-mm and 7-mm in X and Y directions, respectively, see Figure 4.

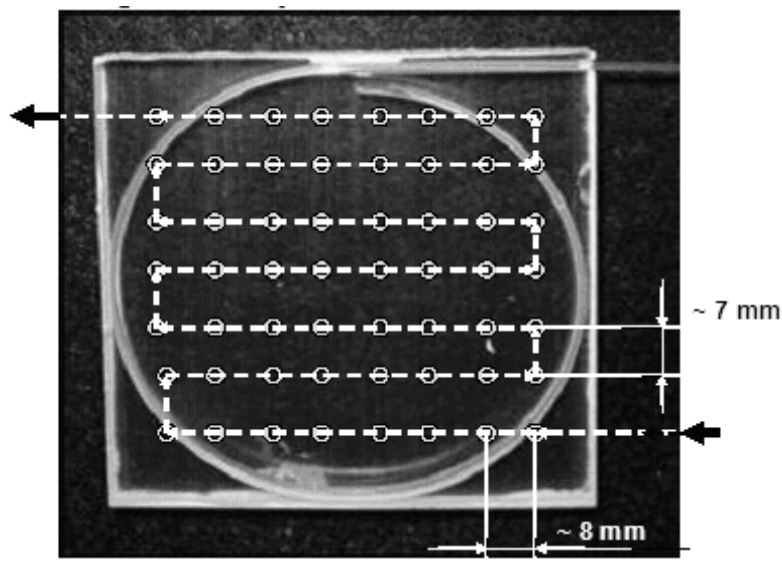

Figure 4 View of scintillating tile with sigma-groove and scheme of two-dimensional scan.

Trigger was made with trigger counter and additional coincidence with signal from tile under test. 4000 events were collected for each point. A Le Croy 2249A ADC was used for signal digitization. Pulse-height distributions collected for each point were fitted with a Gaussian to obtain a peak position. Transverse response maps are presented as a surface plot with a contour, see Figure 5 a), b). Normalization was done on the average response for each tile. The maximum deviation is only inside 5% and 6% for SCSN-81 and FNAL-NICADD samples, respectively. Tile response is highest around boundaries where WLS fibers exit from tile and lowest in the center of tile. Overall, the RMS spread in the transverse uniformity, in both of these cases, is less than 3%. No significant difference between the SCSN-81 and FNAL-NICADD samples were observed.

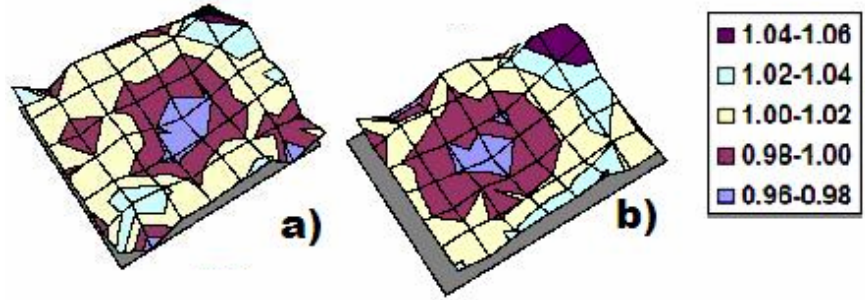

Figure 5. Transverse response variation maps: a) - SCSN-81, b) - FNAL-NICADD

## Conclusion

Our results of tile measurements show, that for relatively small sizes of scintillating tiles with sigma-groove, Kuraray SCSN-81 and FNAL-NICADD extruded scintillating material produces approximately the same results. FNAL-NICADD extruded scintillator would be an acceptable choice, as active media of ALICE Large EMCal.

The final scintillator selection will be based on mechanical and optical tolerances and results of beam studies of calorimeter prototype.